\documentclass[reprint, amsmath, amssymb, aps, superscriptaddress]{revtex4-1}
\usepackage{format}
\usepackage{soul} 
\usepackage{orcidlink}
\usepackage{cleveref}
\allowdisplaybreaks

\begin{document}

\preprint{APS/123-QED}

\title{Universal Threshold Laws for Dipolar Evaporation of Molecular Fermi Gases}

\author{Kevin B. Xu\orcidlink{0000-0002-6929-2852}}
\affiliation{JILA, NIST, and Department of Physics, University of Colorado, Boulder, Colorado 80309, USA}
\author{John L. Bohn\orcidlink{0000-0002-4657-0880}}
\affiliation{JILA, NIST, and Department of Physics, University of Colorado, Boulder, Colorado 80309, USA}
\author{Reuben R. W. Wang\orcidlink{0000-0002-1069-9746}}
\affiliation{ ITAMP, Center for Astrophysics $|$ Harvard \& Smithsonian, Cambridge, Massachusetts 02138, USA }
\affiliation{ Department of Physics, Harvard University, Cambridge, Massachusetts 02138, USA }

\date{\today} 

\begin{abstract}

Fermi statistics suppress the elastic collisions that rethermalize a Fermi gas during evaporative cooling, creating a fundamental low-temperature bottleneck for producing quantum degenerate gases. We show that, for shielded single-component ultracold dipolar molecular Fermi gases in the Wigner threshold regime, this bottleneck is governed by universal threshold laws fixed by the effective dimensionality of the gas.
In this regime, Fermi suppression of rethermalization scales quadratically with temperature in 3D but only linearly in 2D. 
Consequently, the thermalization-to-loss ratio is more favorable in 2D than in 3D, and
these results suggest that 2D geometries are better suited for reaching deep degeneracy. We also present ab initio close-coupling calculations that agree well with experimental measurements on Fermi degenerate gases of KRb molecules, supporting our conclusions.   


\end{abstract}

\maketitle

\textit{Introduction} --
Evaporative cooling has long been the principal route to achieving bulk quantum degenerate gases. For dipolar gases, quantum degeneracy has now been achieved through evaporative cooling of magnetic atoms \cite{Griesmaier05_PRL, Lu11_PRL, Lu12_PRL, Aikawa12_PRL, Aikawa14_PRL1, Miyazawa22_PRL}, and more recently, diatomic polar molecules \cite{Valtolina20_Nat, Schindewolf22_Nat, Bigagli24_Nat, Shi25_arxiv, Lin26_arxiv}. 
Realization of the latter has crucially relied on collisional shielding, which suppresses short-range chemistry \cite{Hu19_Sci} or sticky collisions \cite{mayle_PhysRevA.85.062712, Croft_2017, nichols_PhysRevX.12.011049, Liu_nature, Gersema_2021, bause_prr}. 
Yet shielding alone does not guarantee efficient cooling at deeper degeneracy, as evaporation requires elastic collisions to rethermalize the gas as energetic molecules are gradually removed.

Dipolar Fermi gases are favorable in this respect because their anisotropic and tunable interactions can support efficient elastic rethermalization even in quantum degenerate samples \cite{Li21_Nat, Wang21_PRA, Wang24_PRR, Schindewolf22_Nat, Lin26_arxiv}. At sufficiently low temperature, however, a gas of single-component fermions will always eventually experience a suppression of this rethermalization due to Pauli blocking \cite{Aikawa14_PRL}: as the Fermi sea fills, the available final-state phase space for elastic collisions becomes limited, resulting in a \textit{Fermi suppression} of rethermalization. With modern shielding techniques, molecular Fermi gases have now been cooled to such deep degeneracy that further evaporative cooling is limited by this fundamental Fermi suppression \cite{Lin26_arxiv}. To reach even lower temperatures, it is therefore essential to understand how the competition between rethermalization and loss evolves toward deeper degeneracy.

The effective dimensionality---whether the gas behaves as a three-dimensional (3D) or two-dimensional (2D) system---significantly affects this competition. First, collisional properties, especially scaling laws in the threshold regime, depend on the dimensionality of collisions \cite{Sadeghpour00_JPB}. 
Second, the spatial dimension of the trapped gas informs which shielding techniques are appropriate. In 3D, collisional shielding is achieved either by applying microwaves \cite{Karman18_PRL, Lassabliere18_PRL, Karman25_PRXQ}, a static electric field \cite{Avdeenkov02_PRA, Avdeenkov06_PRA, Gonzalez17_PRA}, or both \cite{Wang26_arxiv, Ho26_PRR, Wang26_inprep}, generating repulsive long-range barriers through field-dressed molecular interactions. In 2D, polarizing ground-state molecules perpendicular to the plane produces predominantly repulsive side-by-side interactions \cite{Micheli07_PRA, Micheli10_PRL, Quemener10_PRA, Quemener11_PRA, Wang25_PRA}. As a result, 3D and 2D gases can exhibit different temperature scalings of thermalization and loss, reflecting differences in density of states, collisional threshold laws, and available shielding mechanisms.


These considerations motivate asking whether evaporative cooling is fundamentally more efficient in 3D or 2D. 
To that end, we show in this work that dimensionality sets universal low-temperature threshold laws for evaporative cooling of single-component dipolar Fermi gases. Specifically, we find scaling laws for both rethermalization and loss in the Wigner threshold regime, which are independent of the shielding technique applied, and show that 2D geometries lead to a slower degradation of evaporative-cooling efficiency at deep degeneracy. We support these scaling laws with ab initio close-coupling calculations for shielded KRb and NaK, which agree well with recent measurements from the JILA KRb experiment \cite{Lin26_arxiv}. 

\textit{Universal Fermi suppression of rethermalization} --
We first isolate the effect of Fermi suppression on rethermalization by considering a low-temperature single-component gas of $N$ fermionic dipoles, with dipole moments aligned along $\hat{\boldsymbol{z}}$. 
The gas is confined in a crossed optical dipole trap that is approximated as harmonic $V(\boldsymbol{r}) = ({m/2}) \sum_{\nu} \omega_\nu^2 r_\nu^2$ with mass $m$. 
The ratios of the harmonic trapping frequencies $\omega_\nu$ along each axis $\nu = x,y,z$ determine the dimensionality of the gas, with 3D having all frequencies on the same order and 2D requiring $\omega_z \gg \omega_x, \omega_y$. 
Throughout this work, ``2D'' refers to the experimentally accessible quasi-2D regime, where the confinement along $\hat{\boldsymbol{z}}$ is strong enough that molecules occupy only the corresponding motional ground state, but not so strong that collisions are strictly two-dimensional \footnote{Since collisions are not strictly two-dimensional, 
shielding in quasi-2D arises from statistical suppression resulting from the symmetrization conditions of indistinguishable fermions \cite{Quemener12_CR}, rather than fully restricted side-to-side dipolar repulsion.
}. 
The threshold laws considered here are the same in true 2D and quasi-2D, however \cite{Ticknor10_PRA}, so our results apply to both regimes.
Finally, given our focus on ultracold temperatures, we primarily consider collisions that occur in the Wigner threshold regime \cite{Sadeghpour00_JPB}.


To quantify the Fermi suppression of elastic rethermalization, we consider cross-dimensional rethermalization experiments \cite{Monroe93_PRL}: rapidly heating a trapped gas along one axis and then allowing it to relax back to thermal equilibrium through elastic collisions. 
At temperatures well below the Fermi temperature, the suppression of thermalization becomes significant due to the filled Fermi sea.
The reduction in thermalization is quantified through a Fermi suppression factor $0 \leq \eta_F \leq 1$ that compares the true thermalization rate $\gamma$ for fermions to the rate expected for a classical gas \cite{Aikawa14_PRL}:
\begin{align} \label{eq:Fermi_rate_relation}
    \gamma
    &=
    \eta_{F} 
    \gamma_{c} / {\rm N_{col}}.
\end{align} 
Here $\gamma_c = \langle n \sigma_{\rm el} v_r\rangle$ is the standard collision rate, which is an ensemble average over the product of the number density $n$, integral elastic cross section $\sigma_{\rm el}$, and relative collision velocity $v_r$ \cite{Monroe93_PRL, Wang21_PRA}. The term ${\rm N_{col}}$ is the number of collisions per rethermalization, which together with the standard collision rate gives the Maxwell-Boltzmann rethermalization rate $\gamma_c/{\rm  N_{col}}$ \footnote{ 
In Ref.~\cite{Zhu13_PRA}, each term in the ratio $\gamma_c/\gamma$ was evaluated without the energy dependence of the elastic cross section, which was appropriate for the experimental temperatures of interest above the Wigner scattering threshold regime. Our current definition includes the threshold energy dependence, with our focus on universal threshold laws. 
}. 


Our key result in this section is that $\eta_F$ has different low-temperature behavior in the regime of deep degeneracy, when the temperature is significantly lower than the Fermi temperature $T/T_F \ll 1$. In 3D, $\eta_{F,\rm{3D}} \propto (T/T_F)^2$ whereas in quasi-2D $\eta_{F,\rm{2D}} \propto T/T_F$.
Thus the elastic rethermalization needed for evaporation is less suppressed in quasi-2D than in 3D when the gas is deeply degenerate. 

We obtain these powers by directly calculating the rethermalization rate $\gamma$ using the Boltzmann equation. Rethermalization is quantified by the relaxation of a variational quantity $\Phi$, chosen here as the deviation of the kinetic energy along the initially excited direction from its isotropic value, $\Phi = p_i^2 - \boldsymbol{p}^2/{\rm d}$, where $\boldsymbol{p}$ is the molecular momentum and the symbol ${\rm d}$ denotes the number of coordinate dimensions.
In our calculations, we assume the initial excitation is $i=z$ in the 3D case and $i=x$ in the 2D case.
The rethermalization rate is then obtained using a variational approach \cite{Massignan05_PRA, Aikawa14_PRL}:
\begin{align} \label{eq:variational_thermalization_rate}
    \gamma
    &\approx 
    -\frac{ \langle {\cal I}[\Phi] \Phi \rangle_F }{ 2 \langle \Phi^2 \rangle_F },
\end{align}
where $\langle \ldots \rangle_F$ denotes a phase-space average incorporating Fermi statistics. It is given by $\langle \ldots \rangle_F = \int d^{\rm d}\boldsymbol{r} d^{\rm d}\boldsymbol{p} f(\boldsymbol{r}, \boldsymbol{p}) [ 1 - h^{\rm d} f(\boldsymbol{r}, \boldsymbol{p}) ] ( \ldots )$, where $f(\boldsymbol{r}, \boldsymbol{p})$ is the phase space distribution and $h$ denotes Planck's constant.
Above, ${\cal I}[\Phi] = {\cal C}[\Phi] / [f ( 1 - h^{\rm d} f)]$ is defined in terms of the collision integral \cite{Reif09_Waveland}
\begin{align} \label{eq:collision_integral}
    {\cal C}[\Phi]
    &=
    \int \frac{ d^{\rm d}\boldsymbol{p}_1 }{ m }
    | \boldsymbol{p} - \boldsymbol{p}_1 |
    f(\boldsymbol{r}, \boldsymbol{p}) f(\boldsymbol{r}, \boldsymbol{p}_1) \nonumber\\
    &\quad\quad \times
    \int d\Omega'
    \frac{ d\sigma_{\rm el} }{ d\Omega' }
    ( 1 - h^{\rm d} f' ) ( 1 - h^{\rm d} f'_1 )
    \Delta\Phi,
\end{align}
where $d\sigma_{\rm el} / d\Omega'$ is the elastic differential cross section with the $({\rm d}-1)$-dimensional solid angle of post-collision relative momentum coordinates $d\Omega'$. The factors $(1 - h^{\rm d} f)$ account for Pauli blocking of final states.
In the limit of small initial excitation, $\Delta \Phi / \Phi \ll1$, the scattering anisotropies have only a weak effect on $\eta_F$ since the distribution remains mostly isotropic in $\boldsymbol{p}$. As such, we focus on rethermalization where the gas is initially excited along a principal axis $i=z$ ($i=x$) in 3D (2D), and the measured thermalization is the rate along that same axis so that $\Phi = p_i^2 - \boldsymbol{p}^2/{\rm d}$. 


For a classical gas, both $\gamma_c$ and ${\rm N_{col}}$ take on universal forms in the threshold regime. The elastic cross sections follow the threshold laws $\sigma_{\rm el} \sim \smash{E_c^0}$ in 3D \cite{Bohn09_NJP} and $\sigma_{\rm el} \sim \smash{E_c^{1/2}}$ in 2D \cite{Ticknor09_PRA}, given the collision energy $E_c = \hbar^2 k^2 / m$.  
Notably, the 2D elastic cross section has an energy dependence that must be included in the integrals of Eq.~(\ref{eq:variational_thermalization_rate}). 
With the threshold elastic cross sections in 3D and 2D previously derived to be $\sigma_{{\rm el}, {\rm 3D}} = 32\pi a_d^2/15$ \cite{Bohn14_PRA}, and $\sigma_{{\rm el}, {\rm 2D}} = 8\pi (\pi - 2) k a_d^2$ \cite{Ticknor09_PRA}, the corresponding collision rates for a classical gas of dipoles at threshold are computed to be $\gamma_{c, {\rm 3D}} = (128 \sqrt{\pi} / 15) \langle n \rangle a_d^2 \sqrt{ k_B T / m }$ and $\gamma_{c, {\rm 2D}} = 16 \pi (\pi - 2) \langle n \rangle a_d^2 (k_B T / \hbar)$.
Above, $a_d = m d^2/(8\pi \epsilon_0 \hbar^2)$ is the dipole length, with $d$ the effective lab-frame dipole moment. 
The number of collisions per rethermalization for $p$-wave dipolar collisions in 2D is found to be ${\rm N_{col}} = 8/3$ (details in App.~\ref{sm:2D} of the Supplementary Material), while the anisotropic expressions for ${\rm N_{col}}$ in a 3D dipolar gas are tabulated in Ref.~\cite{Wang21_PRA}.

\begin{figure}[ht]
    \centering
    \includegraphics[width=\linewidth]{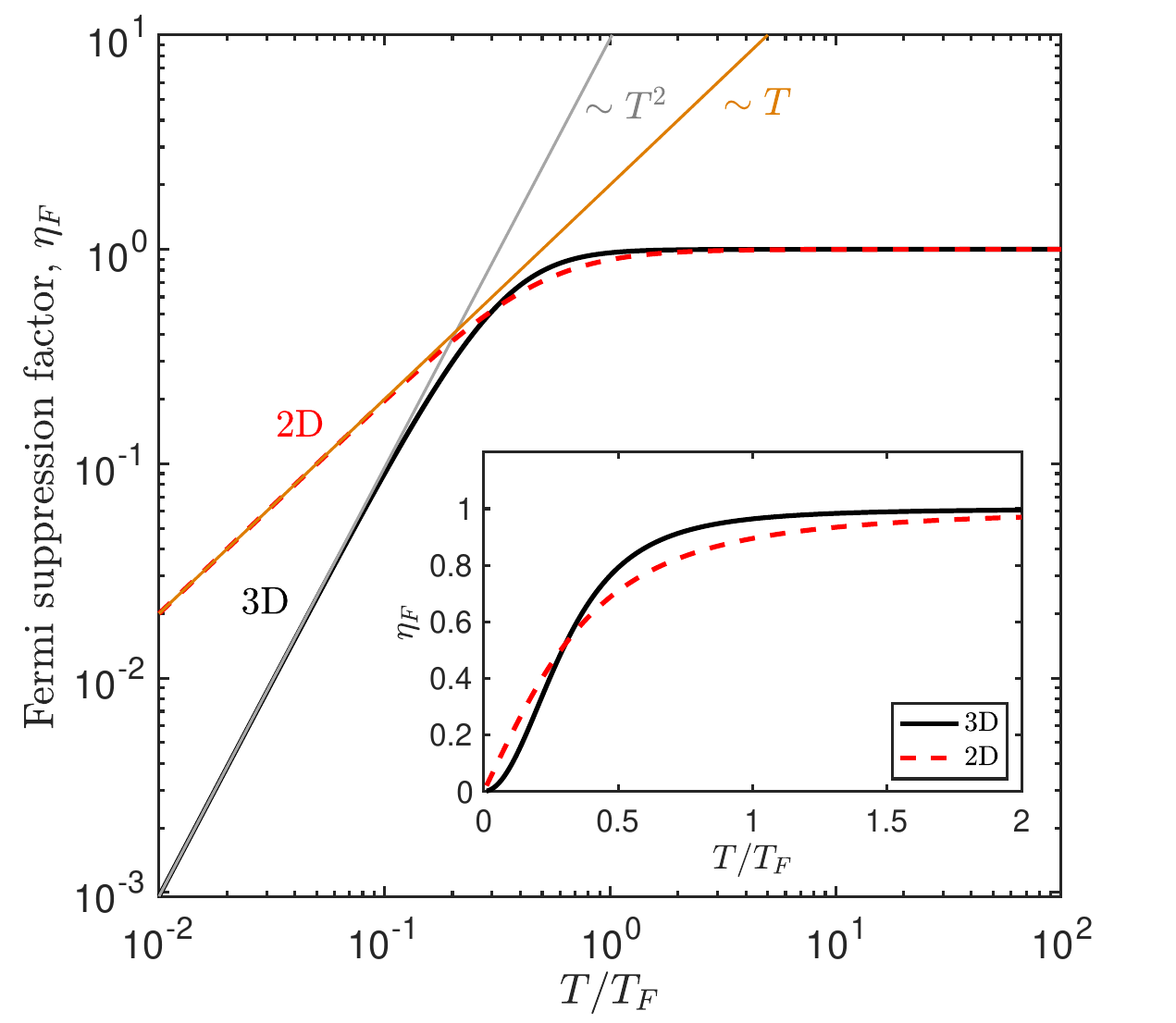}
    \caption{ The Fermi suppression factor $\eta_F$ defined in (\ref{eq:Fermi_rate_relation}) as a function of $T/T_F$ in both 3D (black curve) and 2D (red curve). 
    The thin solid lines show the respective asymptotic scaling of $\eta_F$ for $T/T_F \rightarrow 0$. 
    The inset is a zoomed-in version of the Fermi suppression plot with linear scale axes. The curves shown here use the Wigner-threshold laws (see App.~\ref{app:experiment_vs_theory}).
    }
    \label{fig:Pauli_suppression}
\end{figure}

With these results for the classical rethermalization rate in hand, we compute the Fermi suppression factor via numerical evaluation of Eq.~(\ref{eq:variational_thermalization_rate}) and the relation in (\ref{eq:Fermi_rate_relation}); see Apps.~\ref{sm:thermodynamics}-\ref{sm:2D} of the Supplemental Material for details. The relation for $\eta_F$ as a function of $T/T_F$ is plotted in Fig.~\ref{fig:Pauli_suppression} for both 3D (solid black curve) and 2D (dashed red curve) gases. 
The result exhibits two regimes. For weak to moderate degeneracy, $T/T_F \gtrsim 0.3$, thermalization for the 2D gas is more strongly suppressed than for the 3D gas; however, this trend reverses at deeper degeneracy.
At temperatures $T/T_F \lesssim 0.3$, $\eta_F$ exhibits universal power-law scaling,
falling quadratically with $T/T_F$ in 3D but only linearly in 2D \footnote{ The $\eta_F \sim T^2$ scaling in 3D was also derived and reported in Ref.~\cite{OHara01_PRA}}. This means that once the gas is sufficiently degenerate, the elastic rethermalization needed for evaporation is suppressed more rapidly in 3D than in 2D. Physically, the different exponents are tied to the density of states of a harmonically trapped gas (quadratic in 3D and linear in 2D), rather than to microscopic details of the differential scattering cross section.

\textit{Implications for evaporative cooling} --
Fermi suppression of rethermalization alone does not determine the efficiency of evaporative cooling; rethermalization must remain sufficiently rapid compared with loss. We therefore characterize the competition between these processes by the thermalization-to-loss ratio $\xi = { \gamma / \gamma_L }$ \cite{Ketterle96_AAMOP},
for rethermalization rate $\gamma$ defined in Eq.~(\ref{eq:variational_thermalization_rate}) and loss rate $\gamma_L = \langle n \sigma_{\rm L} v_r \rangle$ with loss cross section $\sigma_{\rm L}$. A larger $\xi$ favors efficient evaporation, while cooling becomes inefficient once molecules are lost faster than the gas can rethermalize. 
For collisionally shielded molecules, $\xi$ can be large, but it remains finite and decreases with decreasing temperature. As such, it is important to quantify the temperature dependencies of both elastic and loss rates for achieving deep degeneracy. 

We now show that the thermalization-to-loss ratio has universal scaling in the deeply degenerate threshold regime. In both dimensions, $\xi$ decreases as the temperature is lowered, but with different powers. The ratio scales as $\xi_{\rm 3D} \propto T^{5/2}$ in 3D, but $\xi_{\rm 2D} \propto T^{2}$ in 2D. 
Thus evaporative cooling eventually becomes inefficient as $T\rightarrow 0$, although the approach to this limit is slower in 2D than 3D. 

The distinct scaling of $\xi$ in 3D and 2D follows from the combined temperature dependence of rethermalization and loss. Having established the former, we now turn to the loss rate $\gamma_L$. To characterize $\gamma_L$, we focus on two representative molecular species, KRb and NaK, that fall into two 3D collisional shielding classes, ${\rm I}$ and ${\rm II}$ \cite{Gonzalez17_PRA}. The class ${\rm I}$ molecule KRb has a relatively small molecular frame dipole moment of $d_0 = 0.57$ D and electronically dominated dispersive interactions \cite{Lepers13_PRA}. Therefore, KRb and other molecules in this class, such as NaLi, have a lower achievable adiabatic shielding potential barrier, and two-body loss for these species results primarily from tunneling through this barrier into the short range. 
By contrast, NaK resides in class ${\rm II}$ with a large molecular frame dipole moment of $d_0 = 2.72$ D, and rotationally dominated long-range interactions. 
Two-body loss in this class of molecules is dominated by long-range inelastic transitions into lower scattering channels. Because both of these losses occur predominantly from scattering on the lowest adiabatic potential that asymptotically connects to the entrance threshold, collisions for fermionic shielded molecules in 3D are $p$-wave dominated \cite{Ni10_Nat}. 
For such ultracold dipolar collisions, both tunneling loss in class I molecules \cite{Bohn09_NJP} and loss from exothermic transitions in class II molecules \cite{Avdeenkov_2005} lead to loss cross sections that scale as $\sigma_L \sim \smash{E_c^{1/2}}$, independent of the shielding technique.



\begin{figure}[ht]
    \centering
    \includegraphics[width=\linewidth]{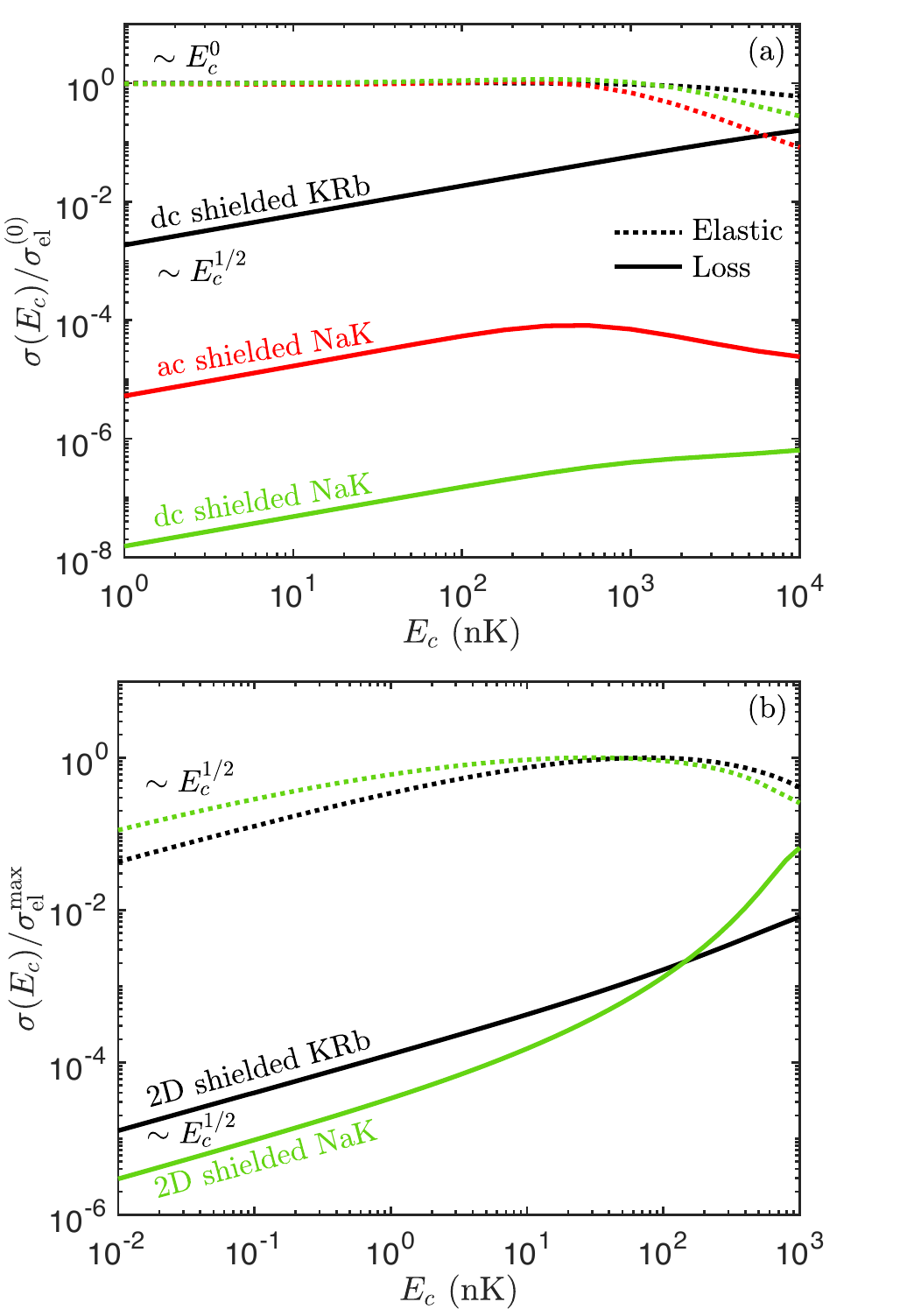}
    \caption{Energy dependence of the elastic (dotted curves) and loss (solid curves) two-body integral cross sections. (a) Cross sections from 3D collisions of dc shielded KRb (black), ac shielded NaK (red), and dc shielded NaK (green). All 3D cross sections are normalized by their respective threshold elastic cross section $\sigma_{\rm el}^{(0)} = 32\pi a_d^2 / 15$. 
    (b) Cross sections from quasi-2D collisions of confinement-induced 2D shielded KRb (black) and dc shielded NaK (green). 
    All 2D cross sections are normalized by the maximum elastic cross section within the plotted energy window. }
    \label{fig:scattering_threshold_comparison}
\end{figure}

To verify the threshold universality of the two-body loss and elastic cross sections in shielded molecules, we perform close-coupling calculations for static field shielded KRb, static field shielded NaK, and microwave shielded NaK. 
We solve the close-coupling equations by numerical log-derivative propagation with a universal loss short-range boundary condition \cite{Wang15_NJP}, using an adaptive step-size Johnson method \cite{Johnson73_JCP}. Convergence of our calculations is ensured by using up to $\ell=9$ collisional partial waves. The shielding parameters used are chosen for highly effective loss suppression. For microwave shielded NaK, we use a Rabi frequency $\Omega = 2\pi \times 20$ MHz and microwave detuning $\Delta = 2\pi \times 10$ MHz, while static field shielding of both KRb and NaK is achieved with an electric field near the F\"orster resonant point
$d_0{\rm E}_{\rm dc} \approx 3.295 B_0$, for rotational constant $B_0$. Fig.~\ref{fig:scattering_threshold_comparison} shows that the close-coupling calculations are in agreement with the predicted threshold laws for both elastic and loss cross sections once the collisions enter the threshold regime. The threshold regime occurs at lower temperatures for NaK than for KRb due to the larger dipole moment of NaK \cite{Bohn09_NJP}.

The same argument for $p$-wave dominated loss applies in 2D, similarly resulting in $\sigma_{L} \sim \smash{E_c^{1/2}}$. 
We confirm this dependence using close-coupling calculations of confinement-induced electric field shielded molecules. Shielding is achieved with an electric field strength of $d_0{\rm E}_{\rm dc} \approx 1.66B_0$, chosen to match the experimental conditions in \cite{Lin26_arxiv}. Quasi-2D confinement is achieved with a trapping frequency of $\omega_z/(2\pi) = 15$ kHz. Numerical propagation is performed with a diabatic-by-sector algorithm as described in the Supplementary Material of Ref.~\cite{Lin26_arxiv}. The results of the close-coupling calculations in Fig.~\ref{fig:scattering_threshold_comparison}(b) confirm that the elastic cross section and loss cross section in 2D both have the threshold scaling $\smash{E_c^{1/2}}$, despite arising from different processes.
 

With these threshold laws in mind, we return to the ratio $\xi$. It is useful to express the in-medium rates $\gamma$ and $\gamma_L$ in terms of effective ensemble-averaged cross sections, rather than direct rates. These effective cross sections remove the explicit density dependence while retaining the influence of Fermi statistics. For elastic collisions, this quantity is obtained from the cross-dimensional rethermalization rate,
from which an effective elastic cross section is extracted as $\sigma_{\rm eff}^{\rm el} = {\rm N}_{\rm col} \gamma / \langle n v_r \rangle = \eta_F \gamma_c / \langle n v_r \rangle$. 
For lossy collisions, the corresponding quantity is obtained from the two-body loss coefficient $\beta_L$ measured through number decay, which gives an analogous effective loss cross section $\sigma_{\rm eff}^{L} = \langle n \rangle \beta_L / \langle n v_r \rangle$. 
Noting that $\gamma_L = \langle n \rangle \beta_L$, the ratio $\xi$ is then straightforwardly given by ${\rm N}_{\rm col}^{-1} \sigma_{\rm eff}^{\rm el}/ \sigma_{\rm eff}^{L}$.

Fermi statistics affect these two effective cross sections in qualitatively different ways. For elastic rethermalization, Fermi suppression drives $\sigma_{\rm eff}^{\rm el} \rightarrow 0$ as $T \rightarrow 0$. However, lossy collisions are not suppressed in the same way, because the final particles either occupy chemically distinct channels or, in the case of inelastic scattering, acquire large momenta outside the occupied Fermi sea.
Instead, the effective loss cross section experiences a saturation due to Fermi statistics. This saturation is an in-medium effect: the bare two-body loss cross section still follows its threshold law, but the ensemble-averaged relative momentum remains finite even at zero temperature because of the finite Fermi momentum in a filled Fermi sea. 
Therefore, although the two-body threshold laws for $\sigma_{L}$ appear favorable for evaporative cooling, the thermalization-to-loss ratio $\xi$ is driven to zero as the temperature decreases, setting a limit where loss begins to dominate and evaporation becomes inefficient.

\begin{figure}
    \centering
    \includegraphics[width=\linewidth]{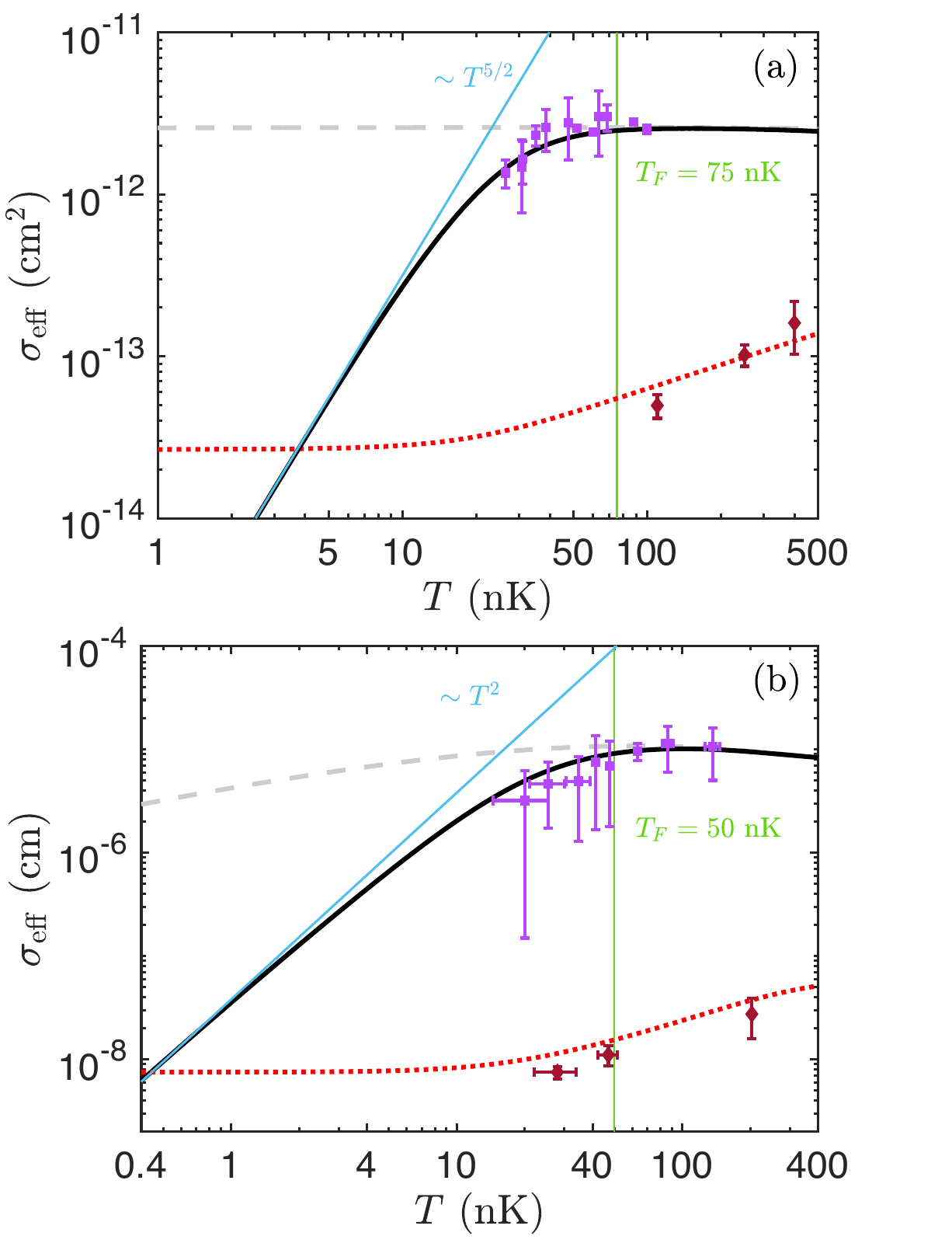}
    \caption{ Comparison of the theory curves to experimental data (points with error bars) from the JILA KRb experiment. (a) Effective 3D integral cross sections for elastic (solid black curve and purple squares) and lossy (dotted red curve and dark red diamonds) collisions. The Fermi temperature of the 3D gas is $T_F = 75$ nK (vertical green line). (b) Effective 2D integral cross sections for elastic (solid black curve and purple squares) and lossy (dotted red curve and dark red diamonds) collisions. In both subplots, the classical effective elastic cross section without Fermi suppression is plotted as the dashed light gray curve. The Fermi temperature of the 2D gas is $T_F = 50$ nK (vertical green line). 
    The respective universal threshold laws are plotted as the thin blue lines, labeled by their temperature scaling. }
    \label{fig:effective_cross_sections_2Dn3D}
\end{figure}

To benchmark this in-medium description, we compute the ensemble-averaged effective cross sections for dc shielded KRb under the experimental conditions of Ref.~\cite{Lin26_arxiv}, in both 3D and 2D, plotted in Fig.~\ref{fig:effective_cross_sections_2Dn3D}. In both dimensionalities, the effective elastic cross sections (solid black curves) are Fermi suppressed from their bare two-body prediction (dashed gray curves) as the gas becomes degenerate. Meanwhile, the effective loss cross sections (dotted red curves) saturate at low temperatures, in contrast to the bare two-body loss cross sections.
Our calculations are also compared with data (points with error bars) from the JILA KRb experiment \cite{Lin26_arxiv}, where these effects were probed for the first time in a molecular quantum gas. Although the 2D rethermalization measurements in Ref.~\cite{Lin26_arxiv} were performed at temperatures outside the threshold regime for KRb, the data are nevertheless consistent with our predictions; see App.~\ref{app:experiment_vs_theory} for a discussion of Fermi suppression in relation to Ref.~\cite{Lin26_arxiv}.


Finally, combining these ingredients gives the low-temperature scaling of the thermalization-to-loss ratio. In 3D, the loss rate saturates, the standard collision rate coefficient scales as $\sqrt{T}$, and the Fermi suppression falls quadratically, giving $\xi_{\rm 3D} \propto T^{5/2}$. In 2D, the loss also saturates, so the weaker Fermi suppression together with the 2D collision rate, both of which scale linearly with $T$, results in $\xi_{\rm 2D} \propto T^{2}$. Thus, $\xi$ vanishes as $T \rightarrow 0$, so evaporation must eventually become inefficient in either geometry; however, the approach to this limit is slower in 2D. As an example, we take the intersection of the KRb $\sigma_{\rm eff}^{\rm el}$ and $\sigma_{\rm eff}^{L}$ curves in Fig.~\ref{fig:effective_cross_sections_2Dn3D} as a proxy for $\xi \sim 1$, corresponding to the onset of inefficient evaporative cooling; this gives approximately 3.7 nK in 3D and 440 pK in 2D.
By this metric, 2D geometries should therefore allow evaporative cooling to reach lower temperatures and deeper Fermi degeneracy than 3D geometries.
Although the use of various molecular species and shielding techniques may change the absolute value of $\xi$, the threshold scalings are universal for single-component dipolar fermions in the Wigner threshold regime and are the main results of this work. 



\textit{Outlook and Conclusions} --
In summary, we have identified universal threshold laws governing the low-temperature evaporative cooling limit of single-component dipolar Fermi gases in 3D and 2D. In the deeply degenerate Wigner threshold regime, Fermi suppression reduces elastic rethermalization as $\eta_F \propto (T/T_F)^{2}$ in 3D but only $\eta_F \propto T/T_F$ in 2D. When loss is included, the thermalization-to-loss ratio decreases as $\xi\propto T^{5/2}$ in 3D and $\xi\propto T^{2}$ in 2D.

These results show that the thermalization-to-loss ratio always vanishes as $T\rightarrow 0$, which means evaporative cooling eventually becomes inefficient as the temperature decreases. This occurs because elastic rethermalization is Fermi suppressed, whereas the loss rate remains nonzero due to the nonzero Fermi momentum. However, dimensionality dictates how rapidly this cooling limit is approached. The thermalization-to-loss ratio decreases more slowly with temperature in 2D, which suggests that 2D geometries may be better suited for reaching deeper Fermi degeneracy. 
Although the magnitude of the thermalization-to-loss ratio can still vary substantially with molecular species and shielding scheme, the low-temperature exponents are fixed by dimensionality within the dipolar threshold regime.

Several extensions to this work remain to be explored. 
Future studies should carefully account for the interplay between temperature and molecule number (and therefore $T_F$) inherent to realistic evaporative cooling protocols, which is not captured in the present treatment. Furthermore, we have neglected Hartree-Fock dipolar mean-field effects, which can cause deformations in the Fermi surface \cite{Miyakawa08_PRA, Sogo09_NJP, Zhang09_PRA, Baillie10_PRA, Biswas26_arxiv}. 
However, these mean-field terms do not directly affect thermalization; rather, they enter indirectly through modifications to the phase space profile, so we expect their inclusion to have a minor effect on the results presented here. 
The gas is also assumed to be dilute throughout this work, such that hydrodynamic effects \cite{Wang23_PRA, Wang23_PRA2} do not play a critical role in evaporation. 
Finally, it would be useful to consider proposed techniques for mitigating Fermi suppression, such as spin mixtures or state-dependent evaporation. Addressing these effects will help determine how the universal dimensional scaling identified here manifests in realistic routes toward deeper Fermi degeneracy.


\textit{Acknowledgments} --
We thank the JILA KRb team for many stimulating discussions and for sharing their experimental data with us.  
KBX and JLB acknowledge support from NSF grant PHY-2408636 and the JILA Physics Frontier Center, PHY-2317149.
RRWW acknowledges support at ITAMP with funding from the National Science Foundation.


\nocite{*}
\bibliography{main.bib} 

\appendix

\section{ \label{app:experiment_vs_theory} Threshold considerations for Fermi suppression}

The 2D Fermi suppression factors shown in Fig.~\ref{fig:Pauli_suppression} differ slightly from those used in Ref.~\cite{Lin26_arxiv}, including Fig.~1E therein. This difference arises from the different collisional regimes considered. The KRb measurements in Ref.~\cite{Lin26_arxiv} for 2D were performed outside the Wigner threshold regime, over a temperature range for which the elastic cross section is approximately constant. 
Accordingly, the theoretical Fermi-suppressed effective cross sections utilized for comparison with experimental measurements were calculated using a constant elastic cross section, which more accurately represents the experimental regime. Concretely, the differential elastic cross section employed in the integral of Eq.~(\ref{eq:collision_integral}) was independent of collision energy. In the present work, our goal is instead to determine the universal low-temperature behavior in the Wigner threshold regime, and we therefore evaluate the suppression using the corresponding threshold scattering forms, for which \(d\sigma_{\rm el}/d\Omega'\propto p_r\) in 2D, where \(p_r=|\mathbf p-\mathbf p_1|\) is the relative momentum. The two 2D calculations describe the same Pauli blocking of final states, but differ slightly in the result because of their different elastic scattering energy dependencies.

Furthermore, over the temperature range accessed in Ref.~\cite{Lin26_arxiv}, the Fermi suppression remains in the higher-temperature regime where rethermalization is more strongly suppressed in 2D than in 3D. Thus, although we compare with the experimental data to validate our formalism, the better evaporation efficiency observed in the 2D experiment should not be attributed to the more favorable asymptotic threshold scaling derived in the present work. In fact, for the experimental temperature regime, Fermi suppression alone inhibits evaporation in 2D more than evaporation in 3D. The lower \(T/T_F\) reached experimentally in 2D instead results from the larger absolute value of the thermalization-to-loss ratio in 2D. The more favorable low-temperature scaling in 2D becomes advantageous close-to-threshold and deeper into quantum degeneracy.

\onecolumngrid

\begin{center}
    {\Large \textbf{Supplemental Material: Universal Threshold Laws for Dipolar Evaporation of Molecular Fermi Gases}}
\end{center} 
\vspace{50pt}

\twocolumngrid

\section{ \label{sm:thermodynamics} Thermodynamics of degenerate Fermi gases }

\subsection{ 3D Fermi gases }

If all the trap frequencies are of comparable strength, the thermodynamic gas is most appropriately treated in three dimensions, where its equilibrium phase space and energy distributions follow Fermi-Dirac profiles \cite{Butts97_PRA}:
\begin{subequations}
\begin{align}
    f_{\rm 3D}(\boldsymbol{r}, \boldsymbol{p})
    &=
    h^{-3}
    \left[  
    1 
    +
    \zeta_{\rm loc}^{-1}(\boldsymbol{r})
    e^{ \frac{ \boldsymbol{p}^2 }{ 2 m k_B T } }
    \right]^{-1}, \\
    f_{\rm 3D}(E)
    &=
    \frac{ E^2 }{ 2 ( \hbar\overline{\omega} )^3 }
    \left(
    1 
    +
    e^{ \frac{ E - \mu }{ k_B T } }
    \right)^{-1}
\end{align}
\end{subequations}
where $\bar \omega$ is the geometrically averaged trap frequency and $\zeta_{\rm loc}(\boldsymbol{r}) = \exp[ \mu_{\rm loc}(\boldsymbol{r}) / ( k_B T ) ]$ is the local fugacity with local chemical potential $\mu_{\rm loc}(\boldsymbol{r}) = \mu - V(\boldsymbol{r})$. 
The equilibrium number density and momentum distribution are obtained by integration against $d^3\boldsymbol{p}$ and $d^3\boldsymbol{r}$ respectively, to give
\begin{subequations}
\begin{align}
    n_{\rm 3D}(\boldsymbol{r})
    &=
    -\frac{ 1 }{ \lambda_{\rm dB}^3 }
    {\rm Li}_{3/2}[ -\zeta_{\rm loc}(\boldsymbol{r}) ], \\
    c_{\rm 3D}(\boldsymbol{p})
    &=
    -6 \left( \frac{ T }{ T_F } \right)^3
    \frac{ {\rm Li}_{3/2}\Big( -\zeta e^{-\frac{ p^2 }{ 2 m k_B T }} \Big) }{ ( 2 \pi m k_B T )^{3/2} },
\end{align}
\end{subequations}
where ${\rm Li}_{n}(z)$ is the polylogarithmic function, $\lambda_{\rm dB}$ is the thermal de Broglie wavelength, and $\zeta=\exp[ \mu/ ( k_B T ) ]$ is the fugacity. 
Integration of $n_{\rm 3D}(\boldsymbol{r})$ over $\boldsymbol{r}$ then gives the equation of state:
\begin{align} \label{eq:fugacity_relation3D}
    -{\rm Li}_{3}(-\zeta) 
    &=
    \frac{ 1 }{ 6 }
    \left(
    \frac{ T }{ T_F^{\rm 3D} }
    \right)^{-3},
\end{align}
where $k_B T_F^{\rm 3D} = \hbar\overline{\omega} ( 6 N )^{1/3}$.

\subsection{ 2D Fermi gases }

When $\omega_z \gg \omega_x, \omega_y$ with all particles occupying the lowest energy $z$-mode, the gas then enters a two-dimensional regime with phase space and energy distributions given by
\begin{subequations}
\begin{align}
    f_{\rm 2D}(\boldsymbol{r}, \boldsymbol{p})
    &=
    h^{-2}
    \left(
    1 
    +
    \zeta_{\rm loc}^{-1}(\boldsymbol{r})
    e^{ \frac{ \boldsymbol{p}^2 }{ 2 m k_B T } }
    \right)^{-1}, \\
    f_{\rm 2D}(E)
    &=
    \frac{ E }{ ( \hbar\overline{\omega} )^2 }
    \left(
    1 
    +
    e^{ \frac{ E - \mu }{ k_B T } }
    \right)^{-1},
\end{align}
\end{subequations}
from which we get the equilibrium number density and momentum distributions:
\begin{subequations}
\begin{align}
    n_{\rm 2D}(\boldsymbol{r})
    &=
    -\frac{ 1 }{ \lambda_{\rm dB}^2 }
    {\rm Li}_1[ -\zeta_{\rm loc}(\boldsymbol{r}) ], \\
    c_{\rm 2D}(\boldsymbol{p})
    &=
    -\frac{ N^{-1} }{ ( m \overline{\omega} \lambda_{\rm dB} )^2 }
    {\rm Li}_1\left[ 
    -\zeta e^{ -\boldsymbol{p^2}/(2 m k_B T) } 
    \right].
\end{align}
\end{subequations}
The equation of state is then computed as
\begin{align}
    -{\rm Li}_{2}(-\zeta)
    &=
    \frac{1}{2}
    \left(
    \frac{ T }{ T_F^{\rm 2D} }
    \right)^{-2},
\end{align}
where $k_B T_F^{\rm 2D} = \hbar \overline{\omega} \sqrt{ 2 N }$.

\section{ Degenerate thermalization in 3D \label{sm:3D}}

In Ref.~\cite{Aikawa14_PRL}, it was shown that the Fermi suppression factor toward thermalization is largely independent of the differential scattering anisotropy. 
As such, we emulate anisotropic dipolar scattering with an energy-independent $p$-wave differential cross section 
\begin{align}
    \frac{ d\sigma }{ d\Omega }(\hat{\boldsymbol{p}}_r, \hat{\boldsymbol{p}}'_r)
    &=
    2 a_p^2 
    (\hat{\boldsymbol{p}}_r \cdot \hat{\boldsymbol{p}}'_r)^2,
\end{align}
where $a_p$ is the $p$-wave scattering length. The corresponding integral cross section is then $\overline{\sigma} = (8\pi/3) a_p^2$. 
In the case of anisotropic scattering, the numerator of the thermalization rate is
\begin{align}
    \langle {\cal I}[\Phi] \Phi \rangle
    &\approx 
    \frac{ 1 }{ m }
    \int d^3\boldsymbol{r} 
    d^3\boldsymbol{p} d^3\boldsymbol{p}_1
    f f_1 \nonumber\\
    &\quad\quad\quad \times
    ( 1 - h^3 f ) ( 1 - h^3 f_1 ) \nonumber\\
    &\quad\quad\quad \times
    ( p^2 + p_1^2 - 2 \boldsymbol{p} \cdot \boldsymbol{p}_1 )^{1/2}
    \Phi 
    \int d\Omega'
    \frac{ d\sigma }{ d\Omega' }
    \Delta\Phi, \nonumber\\
    &=
    \frac{ 1 }{ 2 m }
    \int d^3\boldsymbol{r} 
    d^3\boldsymbol{p} d^3\boldsymbol{p}_1
    f f_1 \nonumber\\
    &\quad\quad\quad \times
    ( 1 - h^3 f ) ( 1 - h^3 f_1 ) \nonumber\\
    &\quad\quad\quad \times
    ( p^2 + p_1^2 - 2 \boldsymbol{p} \cdot \boldsymbol{p}_1 )^{1/2}
    \left(
    p_z^2 - \frac{ \boldsymbol{p}^2 }{ 3 }
    \right) \nonumber\\
    &\quad\quad\quad \times 
    \int d\Omega'
    \frac{ d\sigma }{ d\Omega' }
    ( p^{\prime 2}_{r,z} - p^{2}_{r,z} ),
\end{align}
where evaluating the integral over $\Omega'$ gives us
\begin{align}
    \langle {\cal I}[\Phi] \Phi \rangle
    &\approx 
    \frac{ 1 }{ 2 m }
    \int d^3\boldsymbol{r} 
    d^3\boldsymbol{p} d^3\boldsymbol{p}_1
    f f_1 \nonumber\\
    &\quad\quad\quad \times
    ( 1 - h^3 f ) ( 1 - h^3 f_1 ) \nonumber\\
    &\quad\quad\quad \times
    ( p^2 + p_1^2 - 2 \boldsymbol{p} \cdot \boldsymbol{p}_1 )^{3/2}
    \left(
    p_z^2 - \frac{ \boldsymbol{p}^2 }{ 3 }
    \right) \nonumber\\
    &\quad\quad\quad \times 
    \frac{8 \pi a_p^2 }{15 }
    \left(
    1 - 3 \cos^2\theta_r
    \right),
\end{align}
where $\cos^2\theta_r = { ( \boldsymbol{p}\cdot\hat{\boldsymbol{z}} - \boldsymbol{p}_1\cdot\hat{\boldsymbol{z}} )^2 / ( p^2 + p_1^2 - 2\boldsymbol{p}\cdot\boldsymbol{p}_1 ) }$. 
We once more evaluate the integral over $\boldsymbol{p}$ in the frame where $\hat{\boldsymbol{p}} = \hat{\boldsymbol{z}}$, so that the integral above then becomes 
\begin{align}
    \langle {\cal I}[\Phi] \Phi \rangle
    &\approx 
    \frac{ 1 }{ 2 m }
    \int d^3\boldsymbol{r} 
    d^3\boldsymbol{p} d^3\boldsymbol{p}_1
    f f_1 \nonumber\\
    &\quad\quad\quad \times 
    ( 1 - h^3 f ) ( 1 - h^3 f_1 ) \nonumber\\
    &\quad\quad\quad \times 
    ( p^2 + p_1^2 - 2 {p} {p}_1 \cos\alpha )^{3/2}
    \frac{ 2 p^2 }{ 3 } \nonumber\\
    &\quad\quad\quad \times 
    \frac{8 \pi a_p^2 }{15 }
    \left(
    1 
    - 
    3 \frac{ ( p - p_1\cos\alpha )^2 }{ ( p^2 + p_1^2 - 2{p} {p}_1 \cos\alpha ) }
    \right) \nonumber\\
    &=
    \frac{8 \pi a_p^2 }{ 45 m }
    \int d^3\boldsymbol{r} 
    \int 4\pi p^2 d{p} 
    \int p_1^2 d{p}_1
    f f_1 \nonumber\\
    &\quad\quad\quad \times
    ( 1 - h^3 f ) ( 1 - h^3 f_1 ) \nonumber\\
    &\quad\quad\quad \times 
    \int 2\pi \sin\alpha 
    p^2
    ( p^2 + p_1^2 - 2 {p} {p}_1 \cos\alpha )^{3/2} \nonumber\\
    &\quad\quad\quad \times
    \left(
    1 
    - 
    3 \frac{ ( p - p_1\cos\alpha )^2 }{ ( p^2 + p_1^2 - 2{p} {p}_1 \cos\alpha ) }
    \right) \nonumber\\
    &=
    \frac{8 \pi a_p^2 }{ 45 m }
    \int d^3\boldsymbol{r} 
    \int d{p} d{p}_1
    f f_1 \\
    &\quad\quad\quad \times
    ( 1 - h^3 f ) ( 1 - h^3 f_1 ) \nonumber\\
    &\quad\quad\quad \times 
    \frac{ (4 \pi)^2 }{ 35 } p p_1
    \Big[
    | p - p_1|^5 
    \left( 8 p^2 + p_1^2 + 5 p p_1 \right) \nonumber\\
    &\quad\quad\quad\quad
    \quad\quad
    -
    | p + p_1 |^5 
    \left( 8 p^2 + p_1^2 - 5 p p_1 \right)
    \Big], \nonumber
\end{align}
where $\cos\alpha = \hat{\boldsymbol{p}}\cdot\hat{\boldsymbol{p}}_1$.

The denominator on the other hand, is computed as
\begin{align}
    \langle \Phi^2 \rangle
    &=
    \int d^3\boldsymbol{r} d^3\boldsymbol{p} 
    f ( 1 - h^3 f ) 
    \Phi^2 \nonumber\\
    &=
    \int d^3\boldsymbol{r} d^3\boldsymbol{p} 
    f ( 1 - h^3 f ) 
    \left(
    p_z^2 - \frac{ \boldsymbol{p}^2 }{ 3 }
    \right)^2 \nonumber\\
    &=
    \int d^3\boldsymbol{r} 
    \int 4\pi p^2 dp 
    f ( 1 - h^3 f ) 
    \left( \frac{ 2 p^2 }{ 3 } \right)^2 \nonumber\\
    &=
    \frac{ 4 }{ 9 }
    \int d^3\boldsymbol{r}
    \int 4\pi p^2 d{p} 
    f(\boldsymbol{r}, \boldsymbol{p}) 
    [ 1 - h^3 f(\boldsymbol{r}, \boldsymbol{p}) ]
    p^4.
\end{align}
The remaining integrals are performed numerically.

\section{\label{sm:2D} Degenerate thermalization in 2D }

We now repeat the analysis above in 2D. 
The numerator for $\gamma$ can be written in terms of the integral
\begin{align}
    \langle {\cal I}[\Phi] \Phi \rangle
    &=
    \int d^2\boldsymbol{r} 
    \int d^2\boldsymbol{p} d^2\boldsymbol{p}_1
    \frac{ | \boldsymbol{p} - \boldsymbol{p}_1 | }{ m }
    f f_1
    \Phi \nonumber\\
    &\quad \times 
    \int d\phi'
    \frac{ d\sigma }{ d\phi' }
    ( 1 - h^2 f' ) ( 1 - h^2 f'_1 )
    \Delta\Phi \nonumber\\
    &\approx 
    \int \frac{ d^2\boldsymbol{r} }{ m }
    \int d^2\boldsymbol{p} d^2\boldsymbol{p}_1
    f f_1 \\
    &\quad \times 
    ( 1 - h^2 f ) ( 1 - h^2 f_1 ) \nonumber\\
    &\quad \times 
    ( p^2 + p_1^2 - 2 \boldsymbol{p} \cdot \boldsymbol{p}_1 )^{1/2}
    \Phi 
    \int d\phi'_r
    \frac{ d\sigma }{ d\phi'_r }
    \Delta\Phi, \nonumber
\end{align}
where we will now use $\Phi = p_x^2 - \boldsymbol{p}^2/2$, so that the collisions induce the variation given by
\begin{align}
    \Delta\Phi 
    &=
    \Phi' + \Phi'_{1} - \Phi - \Phi_{1} 
    =
    \frac{ 1 }{ 2 }
    ( p^{\prime 2}_{r,x} - p^{2}_{r,x} ).
\end{align}
Utilizing the 2D dipolar scattering threshold law $\sigma \sim k a_d^2$ \cite{Ticknor09_PRA}, and assuming $p$-wave dominated scattering, we employ a differential scattering cross section $d\sigma/d\phi_r = 8 (\pi - 2) k a_d^2 \cos^2\phi_r$ [i.e. $\sigma = 8 \pi (\pi - 2) k a_d^2$], where $\hbar k = p_r/2$ is the collision wavenumber with the convention such that $\boldsymbol{p}_r = \boldsymbol{p}_1 - \boldsymbol{p}_2$. 
We then evaluate the numerator as:
\begin{align}
    \langle {\cal I}[\Phi] \Phi \rangle
    &\approx 
    \frac{ 1 }{ m }
    \int d^2\boldsymbol{r} 
    \int d^2\boldsymbol{p} d^2\boldsymbol{p}_1
    f f_1 \nonumber\\
    &\quad \times 
    ( 1 - h^2 f ) ( 1 - h^2 f_1 ) \nonumber\\
    &\quad \times 
    ( p^2 + p_1^2 - 2 \boldsymbol{p} \cdot \boldsymbol{p}_1 )^{3/2}
    \left(
    p_x^2 
    -
    \frac{ \boldsymbol{p}^2 }{ 2 }
    \right)  \nonumber\\
    &\quad \times
    \pi (\pi - 2) k a_d^2
    \left(
    1
    -
    2 \frac{ ( \boldsymbol{p}\cdot \hat{\boldsymbol{x}} - \boldsymbol{p}_1\cdot \hat{\boldsymbol{x}} )^2 }{ ( \boldsymbol{p} - \boldsymbol{p}_1 )^2 }
    \right) \nonumber\\
    &=
    \frac{ \pi (\pi - 2) a_d^2 }{ m \hbar }
    \int d^2\boldsymbol{r} 
    \int 2\pi p d{p} 
    \int p_1 d{p}_1
    f f_1 \nonumber\\
    &\quad\quad\quad\quad
    \quad\quad \times
    ( 1 - h^2 f ) ( 1 - h^2 f_1 ) 
    \frac{ p^2 }{ 2 } \nonumber\\
    &\quad\quad\quad\quad \times
    \int d\phi_1
    ( p^2 + p_1^2 - 2 {p} {p}_1 \cos\phi_1 )^{2}  \nonumber\\
    &\quad\quad\quad\quad
    \quad\quad \times
    \left(
    \frac{1}{2}
    -
    \frac{ ( p - {p}_1\cos\phi_1 )^2 }{ p^2 + p_1^2 - 2 p p_1 \cos\phi_1 }
    \right) \nonumber\\
    &=
    -\frac{ \pi (\pi - 2) a_d^2 }{ 2 m \hbar }
    \int d^2\boldsymbol{r} 
    \int 2\pi p d{p} 
    \int p_1 d{p}_1 \nonumber\\
    &\quad\quad\quad\quad \times
    f f_1
    ( 1 - h^2 f ) ( 1 - h^2 f_1 )
    p^2 \nonumber\\
    &\quad\quad\quad\quad \times
    \pi p^2 \left( p^2 + 3 p_1^2 \right) \nonumber\\
    &=
    -\frac{ \pi^2 (\pi - 2) a_d^2 }{ 2 m \hbar }
    \int d^2\boldsymbol{r} 
    \int 2\pi p d{p} 
    \int p_1 d{p}_1 \nonumber\\
    &\quad\quad\quad\quad \times
    f f_1
    ( 1 - h^2 f ) ( 1 - h^2 f_1 ) \nonumber\\
    &\quad\quad\quad\quad \times
    p^4 \left( p^2 + 3 p_1^2 \right).
\end{align}
Meanwhile, the denominator is evaluated as
\begin{align}
    \langle \Phi^2 \rangle
    &=
    \int d^2\boldsymbol{r} d^2\boldsymbol{p} 
    f ( 1 - h^2 f ) 
    \Phi^2 \nonumber\\
    &=
    \int d^2\boldsymbol{r} d^2\boldsymbol{p} 
    f ( 1 - h^2 f ) 
    \left(
    p_x^2 - \frac{ \boldsymbol{p}^2 }{ 2 }
    \right)^2 \nonumber\\
    &\approx
    \int d^2\boldsymbol{r} 
    \int 2\pi p dp 
    f ( 1 - h^2 f ) 
    \left( \frac{ p^2 }{ 2 } \right)^2 \nonumber\\
    &=
    \frac{ 1 }{ 4 }
    \int d^2\boldsymbol{r}
    \int 2\pi p d{p} 
    f
    ( 1 - h^2 f)
    p^4.
\end{align}
All remaining integrals are performed numerically. Taking the Maxwell-Boltzmann limit and comparing to \(\gamma_c\) gives \(\rm N_{\rm col}=8/3\).

\subsection{ Differential scattering in quasi-2D dipolar collisions }

\begin{figure}[ht]
    \centering
    \includegraphics[width=\linewidth]{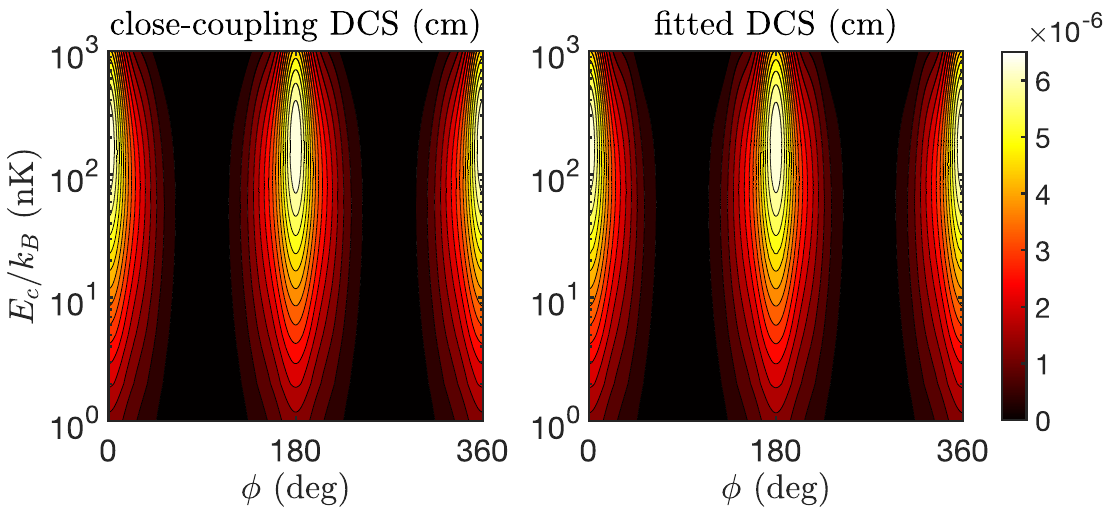}
    \caption{ Differential cross section (DCS) for quasi-2D dipolar collisions as a function of collision energy \(E_c\) and scattering angle \(\phi\). The left panel shows the DCS obtained from close-coupling calculations, while the right panel shows the fit to the angular expansion of Eq.~(\ref{eq:DCS2D_fit}) using \(\ell=1,\ldots,6\). The close agreement validates the fitted representation used to efficiently calculate \(\rm N_{\rm col}\).}
    \label{fig:closecouplingDCS_vs_fittedDCS}
\end{figure}

To compute ${\rm N}_{\rm col}$ from the differential cross section (DCS) obtained from close-coupling calculations in quasi-2D, we utilize the functional form
\begin{align} \label{eq:DCS2D_fit}
    \frac{ d\sigma }{ d\phi'_r }
    &=
    \sigma(p_r)
    \sum_{\ell}
    \lambda_{\ell}(E_c)
    \cos^{2\ell}(\phi'_r - \phi_r),
\end{align}
where $\lambda_{\ell}$ are the angular expansion coefficients that only depend on the collision energy. In practice, we utilize $\ell = 1,2,...,6$.
The collision energy is given by $E_c = \hbar^2 k^2 / (2\mu) = p_r^2 / (8\mu) = ( p^2 + p_1^2 - 2 \boldsymbol{p} \cdot \boldsymbol{p}_1 ) / (8\mu)$. 
Plugging the ansatz above into the integral over the differential cross section gives
\begin{align}
    & \int d\phi'_r
    \frac{ d\sigma }{ d\phi'_r }
    \Delta\Phi \nonumber\\
    =&
    \sigma(p_r)
    \sum_{\ell}
    \lambda_{\ell}(p_r)
    \int d\phi'_r
    \cos^{2\ell}(\phi'_r - \phi_r) 
    \frac{ 1 }{ 2 }
    ( p^{\prime 2}_{r,x} - p^{2}_{r,x} ) \nonumber\\
    =&
    \frac{ \sigma(p_r) }{ 2 }
    p_r^2
    \sum_{\ell}
    \lambda_{\ell}(p_r)
    \int d\phi'_r
    \cos^{2\ell}(\phi'_r - \phi_r) \nonumber\\
    &\quad\quad\quad\quad
    \quad\quad\quad\quad
    \quad\quad \times 
    ( \cos^{2}\phi'_r - \cos^{2}\phi_r ),
\end{align}
for which we tabulate the integrals over $\phi'_r$ for each $\ell$ below:
\begin{align}
    \begin{tabular}{l|c}
    $\ell$ & $\int d\phi'_r
    \cos^{2\ell}(\phi'_r - \phi_r) 
    ( \cos^{2}\phi'_r - \cos^{2}\phi_r )$ \\
    \hline
    1 & $-\frac{ \pi }{ 4 } (2 \cos^2\phi_r - 1)$ \\
    2 & $-\frac{ \pi }{ 8 } (2 \cos^2\phi_r - 1)$ \\
    3 & $-\frac{ 5 \pi }{6 4 } (2 \cos^2\phi_r - 1)$ \\
    4 & $-\frac{ 7 \pi }{ 128 } (2 \cos^2\phi_r - 1)$ \\
    5 & $-\frac{ 21 \pi }{ 512 } (2 \cos^2\phi_r - 1)$ \\
    6 & $-\frac{33 \pi }{ 1024 } (2 \cos^2\phi_r - 1)$
    \end{tabular}.
\end{align}
A comparison of the DCS obtained from close-coupling calculations and the fitted functional form of Eq.~(\ref{eq:DCS2D_fit}) is plotted in the left and right subplots of Fig.~\ref{fig:closecouplingDCS_vs_fittedDCS} respectively, showing excellent agreement.  

Utilizing these closed form expressions allows efficient computation of ${\rm N}_{\rm col}$ in quasi-2D gases of dipolar molecules with accurate differential cross sections.

\end{document}